\documentclass[aps,prl,reprint,superscriptaddress,floatfix]{revtex4-2}
\usepackage{graphicx}
\usepackage{dcolumn}
\usepackage{hyperref}
\usepackage{amsmath}
\usepackage{amssymb}
\usepackage{braket}
\usepackage{upgreek}
\usepackage[version=4]{mhchem}
\usepackage[binary-units]{siunitx}[=v2]
\usepackage{booktabs}
\DeclareSIUnit{\torr}{Torr}

\usepackage[utf8]{inputenc}
\usepackage[T1]{fontenc}

\begin{document}

\title{Optical Memory in a Microfabricated Rubidium Vapor Cell}

\author{Roberto Mottola}
\email{roberto.mottola@unibas.ch}
\author{Gianni Buser}
\author{Philipp Treutlein}
\email{philipp.treutlein@unibas.ch}
\affiliation{Departement Physik, Universit\"{a}t Basel, Klingelbergstr. 82, 4056 Basel, Switzerland.}

\date{December 19, 2023} 

\begin{abstract}
	Scalability presents a central platform challenge for the components of current quantum network implementations that can be addressed by microfabrication techniques. We demonstrate a high-bandwidth optical memory using a warm alkali atom ensemble in a microfabricated vapor cell compatible with wafer-scale fabrication.	
	By applying an external tesla-order magnetic field, we explore a novel ground-state quantum memory scheme in the hyperfine Paschen-Back regime, where individual optical transitions can be addressed in a Doppler-broadened medium. Working on the $^{87}$Rb D$_2$ line, where deterministic quantum dot single-photon sources are available, we demonstrate bandwidth-matching with hundreds of megahertz broad light pulses keeping such sources in mind. For a storage time of \SI{80}{\nano\second} we measure an end-to-end efficiency of $\eta_{e2e}^{\SI{80}{\nano\second}} = \SI{3.12\pm.17}{\percent}$, corresponding to an internal efficiency of $\eta_{\text{int}}^{\SI{0}{\nano\second}} = \SI{24\pm3}{\percent}$, while achieving a signal-to-noise ratio of $\text{SNR} = \num{7.9\pm.8}$ with coherent pulses at the single-photon level.
\end{abstract}

\maketitle

\emph{Introduction.}---Quantum networks \cite{Kimble2008,Wehner2018} are envisioned to enable secure quantum communication \cite{Gisin2002,Scarani2009,Xu2020} and distributed \cite{VanMeter2016} and blind quantum computing \cite{Barz2011}, as well as performing precise measurements through novel forms of distributed quantum sensing \cite{Gottesman2012,Komar2014,Eldredge2018}.
Optical quantum memories and matched single-photon sources are the building blocks of such networks. Direct connections between individual network nodes have been realized \cite{Wei2022}, with complex systems such as cavity-trapped atoms \cite{Noelleke2013} or hybrid interfaces \cite{Maring2017,Puigibert2020} regularly pushing performance records. Remote network operations have also been demonstrated both in cold ensembles \cite{Jing2019} and diamond vacancy centers \cite{Pompili2021, Hermans2022}. As the principle feasibility of quantum networking becomes tangible, new central challenges including scalability, speed, and ease of use in deployment emerge. Platforms compatible with room-temperature operation and mass fabrication must enter any realistic vision of larger networks.

The experimental simplicity of warm alkali vapor memories is attractive for scaling quantum networks \cite{Glorieux2023}. Recently, single-photon storage and retrieval has been demonstrated in broadband vapor-based memories both in ladder \cite{Kaczmarek2018,Davidson2023} and $\Lambda$-schemes \cite{Buser2022}. Moreover, micro-electromechanical system (MEMS) fabrication technology has already successfully miniaturized vapor cells for diverse applications, including quantum sensors such as atomic clocks, gyroscopes, and magnetometers \cite{Kitching2018}. However, a quantum memory has not yet been realized in a MEMS cell. Demands on vapor cell properties vary with application, and key memory demands including sufficient optical path and molecular buffer gas pressures on the order of tens of millibar \cite{Klein2009} are not typical features of microfabricated clock cells. Further, common methods of sealing MEMS cells such as anodic bonding involve temperatures incompatible with spin anti-relaxation coatings on the cell walls \cite{Chi2020}, which are critical to achieving long storage times in memories. Nevertheless, the state-of-the-art in cell fabrication has addressed these issues with techniques including cell-internal light routing, lower temperature bonding methods, and techniques to achieve arbitrary buffer fillings on wafer scales \cite{Bopp2020,Maurice2022,McGilligan2022}.
These developments pave the way for scalability and spatial multiplexing -- a wafer of suitable vapor cells potentially provides hundreds of independent memories -- and could even enable satellite-borne applications \cite{Guendogan2021}.

In this Letter, we present the experimental realization of an optical quantum memory protocol in a microfabricated vapor cell on the $^{87}$Rb D$_2$ line. By applying a static, tesla-order, external magnetic field, degeneracy in the atomic level structure is lifted in such a way that a near-ideal three-level $\Lambda$ system can be addressed. As a proof-of-principle, weak coherent pulses attenuated to the single-photon level are stored and retrieved, with operational bandwidths of hundreds of megahertz. By investigating such fast signals, we demonstrate the applicability both of our novel scheme and of the miniaturized vapor cell to the task of storing single photons as they are produced by high quality solid state emitters -- for instance, GaAs quantum dots constitute excellent single-photon sources emitting at the Rb D$_2$ line \cite{Zhai2022}. Interfacing deterministic photon sources with vapor cell memories would enable network architectures such as the one proposed in Ref.~\cite{Sangouard2007}. Moreover, by evaluating memory performance and limitations, we establish a list of requirements for cell fabrication aimed at memory applications, while simultaneously exploring atomic memory interactions in a hitherto untested physical parameter range.

\begin{figure}
	\begin{center}
	\includegraphics[width=\columnwidth]{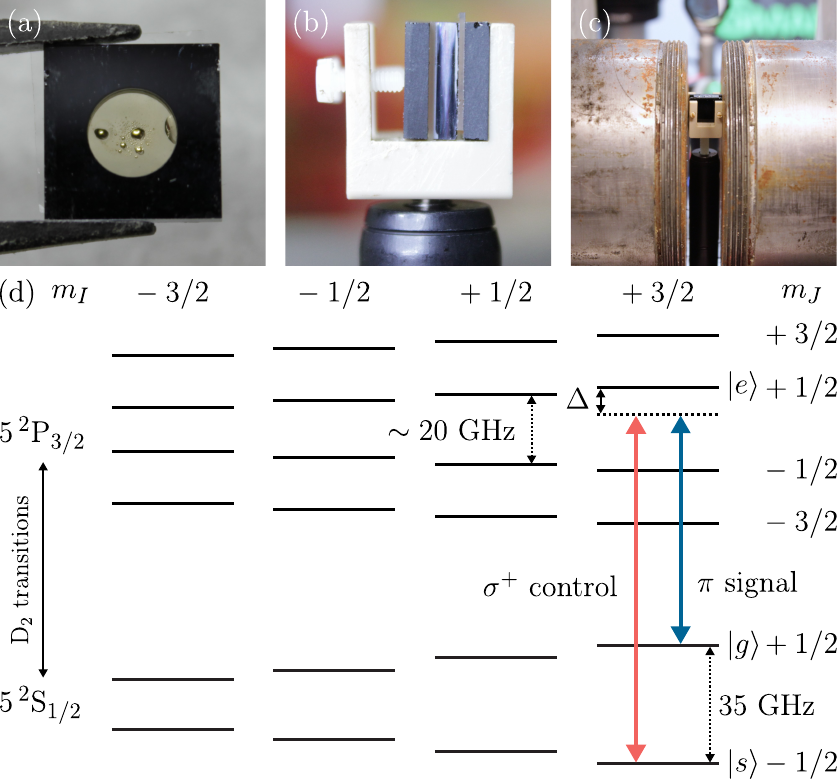}
	\caption{\label{scheme} Front (a) and side (b) view of the microfabricated vapor cell. Colored glass filters for absorbing the heating lasers sandwich the cell. (c) The ferromagnetic cores of the electromagnet limit the physical and optical access close to the cell. (d) Energy levels of $^{87}$Rb in the external field and $\Lambda$-scheme used for the atomic memory. Energy splitting and decoupling of $\mathbf{J}$ and $\mathbf{I}$ yield a ``clean'' three-level system in the Doppler broadened $^{87}$Rb D$_2$ line. The signal and control fields are on two-photon resonance with a detuning $\Delta = -2\pi \times \SI{750}{\mega\hertz}$ with respect to $\ket{e}$.}
	\end{center}
\end{figure}

\emph{Memory scheme and setup.}---The quantum memory protocol is implemented in a microfabricated atomic vapor cell with \SI{2}{\milli\meter} internal thickness and an aperture diameter of \SI{5}{\milli\meter} [see Fig.~\ref{scheme}(a)]. The interaction is based on a ground state $\Lambda$ scheme. Atoms are initially prepared in state $\ket{g}$ by optical pumping. An incoming signal photon is stored by mapping it to a collective spinwave excitation between the two metastable ground states $\ket{g}$ and $\ket{s}$ by a strong control pulse resonant to the second ``leg'' of the $\Lambda$ system. By applying a second control pulse after a variable delay, the signal is retrieved from the memory. 

Generally speaking, the energetic structure of atoms is more complex than just three levels. In order to avoid spurious noise processes such as four-wave mixing, the isolation of a near-ideal three-level system is crucial. To accomplish this goal, we operate the memory in the hyperfine Paschen-Back regime, reached by applying a large, static B-field. The resulting decoupling between nuclear spin manifolds and the large energy level splittings within each manifold caused by the Zeeman effect allow us to obtain the necessary resolution of individual transitions to address a spectrally well-isolated, clean, three-level $\Lambda$-system for storage, as illustrated in Fig.~\ref{scheme}(d).
The magnetic field is applied orthogonally to the propagation axis of the light, a configuration known as Voigt geometry, by a Bruker B-E 10 electromagnet for flexibility in testing [see Fig.~\ref{scheme}(c)]. In a future miniaturized setup, we envision employing a suitably designed permanent magnet akin to what is used in \cite{PoncianoOjeda2020}. For a thorough spectroscopic characterization of this regime, information on the viability of state preparation therein, and a broader overview of its usefulness in establishing quantum control in hot vapors, we point to our companion paper \cite{Mottola2023PRA}.
   
The memory operates on the $^{87}$Rb D$_2$ line at \SI{780}{\nano\meter} in a $\SI{1.06}{\tesla}$ magnetic field. 
The vapor cell is filled with enriched $^{87}$Rb (abundance $\leq \SI{90}{\percent}$) and about \SI{11}{\milli\bar} of Ar buffer gas to impede atomic motion. The fabrication process of such MEMS type cells is described in \cite{DiFrancesco2010,Pellaton2012}. The cell is sandwiched by two \SI{2}{\milli\meter}-thick pieces of RG9 filter glass [see Fig.~\ref{scheme}(b)] and heated with two multimode, telecom lasers in order to increase the Rb vapor pressure and reach moderate optical depths. With this heating technique, we can achieve atomic temperatures $>\SI{130}{\degreeCelsius}$. 

We choose the signal to be near resonant to the $\pi$-transition coupling the stretched state $\ket{g} = \ket{m_J=+\frac{1}{2},m_I=+\frac{3}{2}}$ to the excited state $\ket{e} = \ket{m'_J=+\frac{1}{2},m'_I=+\frac{3}{2}}$ for its higher transition strength. Consequently, the storage state is $\ket{s} = \ket{m_J=-\frac{1}{2},m_I=+\frac{3}{2}}$ and the control field must be $\sigma^+$ polarized. In the experiment, the control field is applied with linear polarization orthogonal to the signal, corresponding to an equal superposition of $\sigma^+$ and  $\sigma^-$ (see \cite{Mottola2023PRA} for details).
Initially, the $m_I=+\frac{3}{2}$-manifold is prepared in $\ket{g}$ through optical pumping. A second pump laser is used to drive the singly forbidden transition $\ket{m_J=+\frac{1}{2},m_I=+\frac{1}{2}} \longleftrightarrow \ket{m'_J=-\frac{3}{2},m'_I=+\frac{3}{2}}$ to partially polarize the nuclear spin and increase the number of atoms in the addressed $m_I$-manifold.
A working point detuning $\Delta = -2\pi \times \SI{750}{\mega\hertz}$ from the excited state $\ket{e}$ is empirically found to optimize the memory performance under experimental operating conditions.

\begin{figure}
	\includegraphics[width=\columnwidth]{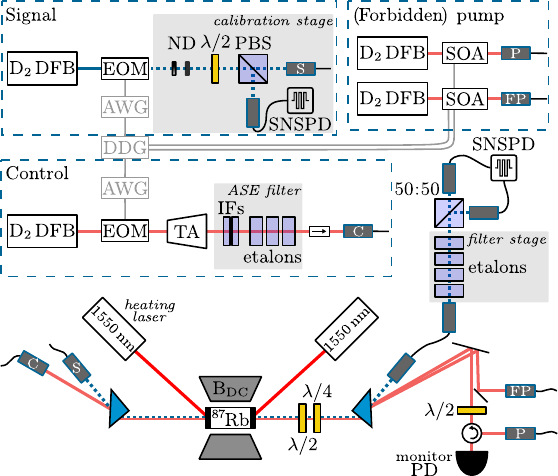}
	\caption{\label{setup} Sketch of the experimental setup. The dashed boxes represent the preparation stages for the various optical pulses involved in the memory protocol. DFB, distributed feedback (laser); EOM, electro-optic modulator; AWG, arbitrary waveform generator; DDG, digital delay generator; SOA, semiconductor optical amplifier; TA, tapered amplifier; IF, interference filter; $\lambda/2$, half-wave plate; $\lambda/4$, quarter-wave plate; 50:50, beam splitter; SNSPD, superconducting nanowire single-photon detector. The labels S, C, P, and FP represent the fiber links of signal, control, pump, and forbidden pump, respectively.}
\end{figure}

\begin{figure}[b]
	\includegraphics[width=\columnwidth]{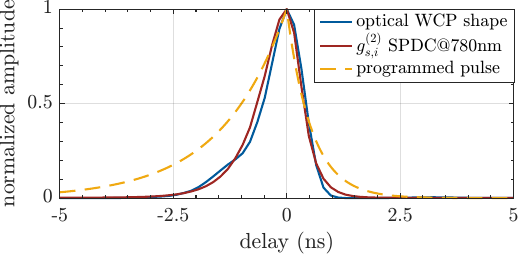}
	\caption{\label{tshape} Signal pulse shape. The blue trace shows the optical weak coherent pulse (WCP), measured as signal input (Fourier bandwidth \SI{0.7}{\giga\hertz} FWHM). The target shape, consisting of the cross-correlation $g^{(2)}_{s,i}$ from the spontaneous parametric downconversion (SPDC) source in \cite{Buser2022}, is shown in red. For comparison, the dashed yellow line shows the voltage pulse programmed in the AWG.}
\end{figure}

The optical setup is sketched in Fig.~\ref{setup}.
Optical pulses are generated from cw DFB lasers by amplitude modulation with EOMs (Jenoptik AM785), with their waveforms set by an AWG (PicoQuant PPG512).
The signal pulses are subsequently attenuated to the single-photon level and their temporal shape, shown in Fig.~\ref{tshape}, is matched to that of heralded photons from a parametric downconversion source described in \cite{Buser2022}.
The Gaussian-shaped control pulses are amplified to the required intensities with a TA. Amplified spontaneous emission (ASE) from the TA is filtered with two interference filters (Laseroptik, \SI{0.37}{\nano\meter} FWHM bandwidth, specified by manufacturer) 
and three monolithic etalons [two \SI{1150\pm20}{\mega\hertz} FWHM bandwidth; one \SI{280\pm10}{\mega\hertz} FWHM bandwidth]. All etalons herein reach suppressions of \SI{-33(1)}{\decibel} at $\text{FSR}/2$.

Signal and control are each coupled into single-mode (SM) fibers and are combined at the memory stage on a polarizing calcite prism. 
The signal (control) is focused with the outcoupler to a $1/e^2$ diameter at the center of the cell of \SI{185\pm6}{\micro\meter} [\SI{440\pm6}{\micro\meter}]. The beam waists are measured at an equivalent distance from the cell center with a beam profiler. The control beam is chosen to be larger for a more homogeneous Rabi frequency across the spatial mode of the signal, which is known to improve the memory efficiency at lower peak $\Omega$ in simulation \cite{Buser2022}.

The atomic ensemble is optically pumped before each storage attempt with a dedicated pump laser tuned to resonance with $\ket{s}\rightarrow\ket{e}$. The pump counter-propagates with the control beam and is coupled in through an optical circulator in the control arm. A semiconductor optical amplifier is used as a fast optical switch to turn off the pumping beam, as described in \cite{Buser2021}. 
The forbidden transition is pumped by a further laser, which is aligned under a small angle [estimated trigonometrically to be \SI{2.6\pm0.9}{\milli\radian}] to the main pump using a D-shaped mirror. Both pump lasers are collimated and have a diameter of approximately \SI{2}{\milli\meter}.
In order to maximize depletion of $\ket{s}$, the pump laser is switched off \SI{50}{\nano\second} later than the forbidden pump.
The optical power on the atoms from the pump and forbidden pump lasers is \SI{14.8\pm0.4}{\milli\watt} and \SI{16.0\pm0.5}{\milli\watt}, respectively.

A second prism filters the retrieved signal by polarization. Over 8 orders of magnitude of control suppression are achieved this way. After coupling the signal into an SM fiber, a stack of monolithic etalons filters spectrally. Three of these are chosen to have an $\text{FSR}=\SI{71.1}{\giga\hertz}$, equal to twice the ground state splitting of $^{87}$Rb in the external field [\SI{1.45}{\milli\meter} thickness, FWHM bandwidth \SI{1.19(2)}{\giga\hertz}], and one [\SI{4}{\milli\meter} thickness, FWHM bandwidth \SI{550\pm10}{\mega\hertz}] matches the bandwidth of the retrieved photon.
Finally, the signal is evenly split, fiber coupled, and detected with SNSPD (Single Quantum Eos).
When the vapor cell is cold, the transmission through the whole setup at the signal frequency is measured to be slightly below \SI{20}{\percent} for a strong cw probe.

Currently, the experiment is repeated periodically and triggered by a digital delay generator (Highland Technology T564) with a repetition rate of $\SI{300}{\kilo\hertz}$, allowing for \SI{2.8}{\micro\second} of optical pumping. Nevertheless, electronic and optical switches are all suitable for asynchronous operation, as is required when paired with a probabilistic single photon source (cf. Ref.~\cite{Buser2022}). 

\emph{Storage and retrieval.}---Memory experiments were performed for a storage time of about \SI{80}{\nano\second} at an atomic temperature $T = \SI{90\pm1}{\degreeCelsius}$, determined spectroscopically. 
Gaussian control pulses with a duration of $\text{FWHM} = \SI{3.8}{\nano\second}$ were used, with a maximum available peak Rabi frequency of $2\pi\times\SI{683(15)}{\mega\hertz}$ at the given waist. 
The signal pulse attenuation was calibrated right before the storage and retrieval measurement. From the counts measured on the calibration channel, within a \SI{6.48}{\nano\second}-wide region of interest (ROI) around the signal pulse, a mean photon number of $\lvert \alpha \rvert ^2 = 0.97(6)$ was set. 
Detection events are tagged with a time-to-digital converter (qutools quTAU). The counts from the Hanbury Brown and Twiss configured detectors after the memory are added for evaluation. All figures of merit are specified using a \SI{6.48}{\nano\second} ROI. 

\begin{figure}
	\includegraphics[width=\columnwidth]{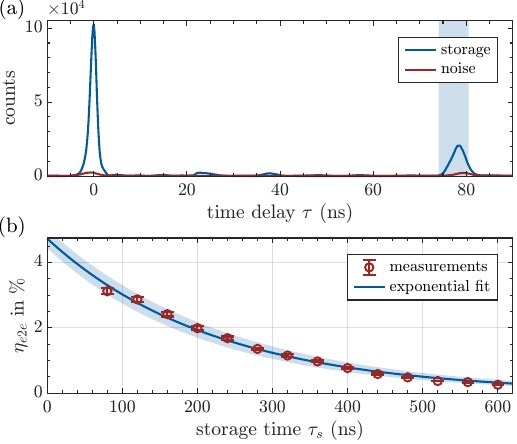}
	\caption{\label{storage} (a) Arrival-time histogram for storage and retrieval experiments and corresponding noise measurements. Zero time delay corresponds to the time at which signal photons leaking through the memory in failed storage attempts reach the detector. The shaded area corresponds to the \SI{6.48}{\nano\second}-wide ROI. The features between \num{20} and \SI{40}{\nano\second} constitute unintentional read-out and are induced by insufficient suppression of the control EOM. (b) Memory lifetime measured as the decrease in efficiency. Each data point is integrated over \SI{2}{\minute} and the data are scaled by a factor \num{1.13} to match the performance of the data set shown in (a). A fitted exponential decay describes the data well. The shaded area corresponds to the \SI{95}{\percent} confidence interval of the fit.}
\end{figure}

A photon arrival-time histogram is shown in Fig.~\ref{storage}(a).
For an integration time of $t_{\text{int}} \approx \SI{1}{\min}$, the memory receives $N_{\text{trig}} = \num{1.81e7}$ triggers, each corresponding to a storage attempt, and $N_{\text{ret}} = \num{4.46e5}$ counts accumulate within the retrieval window.
To estimate noise during read-out, the experiment is repeated with the input physically blocked. Within the same ROI, $\num{4.28e4}$ noise counts are detected. This noise measurement does not account for the finite suppression of the signal EOM. During the retrieval time window, the leakage through the EOM is static. To account for this, we add the averaged difference of \num{183} counts per bin between the storage and noise traces far from signal switching to the noise counts, obtaining $N_{\text{noise}}$. Over the ROI we estimate that this offset adds \num{7325} spurious counts. 
The detected counts correspond to a signal-to-noise ratio of $(N_{\text{ret}} - N_{\text{noise}})/N_{\text{noise}} = \num{7.9\pm.8}$.

From the latter quantity $\mu_1 = \eta_{\text{det}}^{\text{HBT}}/\text{SNR} = \num{0.09\pm0.01}$ can be computed, describing the required mean photon number at the input to reach $\text{SNR} = 1$. Here $\eta_{\text{det}}^{\text{HBT}} = \SI{70\pm4}{\percent}$ is the detection probability of a weak coherent pulse in an HBT setup with non-photon-number-resolving detectors given by  $\eta_{\text{det}}^{\text{HBT}}(\lvert\alpha\rvert^2)=2\left[1-\exp(-\lvert\alpha\sqrt{\eta_{\text{det}}/2}\rvert^2)\right]$, where $\eta_{\text{det}}$ is the detection efficiency of the SNSPD.
The measurements yield an end-to-end efficiency of $\eta_{e2e}^{\SI{80}{\nano\second}} = (N_{\text{ret}} - N_{\text{noise}}) / (\eta_{\text{det}}^{\text{HBT}} N_{\text{trig}}) = \SI{3.12\pm.17}{\percent}$.
In the region between the read-in and read-out control pulses in Fig.~\ref{storage}(a), unintentional retrieval can be observed. This is caused by ringing in the control EOM, which is translated to an optical signal. Taking into account some minor dynamics caused by the signal EOM during the storage time, we estimate these detrimental features to account for a \SI{0.45\pm.05}{\percent} loss in end-to-end efficiency. 

The memory lifetime is measured by the decrease in efficiency over time. This measurement was performed with a repetition rate of $\SI{100}{\kilo\hertz}$, allowing us to use the same pump pattern for all data, guaranteeing unaltered initial conditions. The end-to-end efficiency as a function of storage time is plotted in Fig.~\ref{storage}(b). The lifetime data are scaled by a factor \num{1.13} to match the measured efficiency for a storage time of \SI{80}{\nano\second} to the one obtained from the data set shown in Fig.~\ref{storage}(a). This discrepancy is a question of day-to-day optimization and the correction is applied to accurately capture the internal efficiency of the main result.
Without any direct optimization attempt, a $1/e$ lifetime of $\SI{224\pm8}{\nano\second}$ was achieved. The data are well fit by a decaying exponential (solid line). This functional form indicates the lifetime is limited by loss processes, as dephasing of the spin wave is described by Gaussian decay \cite{Zhao2008}. Although this order of magnitude is predicted, the lifetime is shorter than expected from simple models of transverse atomic motion out of the interaction region \cite{Novikova2011}. A reduction due to motion along the optical axis leading to spin destruction collisions with the uncoated cell windows is plausible.
Correcting for technical losses by dividing out the signal transmission through the setup, the internal efficiency at zero storage time is estimated to be $\eta_{\text{int}}^{\SI{0}{\nano\second}} = \SI{24\pm3}{\percent}$.

\emph{Discussion.}---The microfabricated vapor cell design and its filling were not optimized for memory application, leaving much room for future improvement.
A pump-probe relaxation in the dark measurement shows that an atomic polarization of \SI{88(1)}{\percent} in $\ket{g}$ is reached within the $m_I = +\frac{3}{2}$-manifold \cite{Mottola2023}. Currently, the polarization is most likely limited by radiation trapping due to the high Rb number density $n$ at the temperature required for efficient storage. At \SI{90}{\degreeCelsius} and accounting for \SI{90}{\percent} enrichment, the number density in the considered nuclear spin manifold is $n=\SI{5.5e11}{\per\centi\meter\cubed}$.
A molecular buffer gas, such as N$_2$, would quench the excited state and improve the atomic polarization \cite{Rosenberry2007}. By more fully depleting the storage state in this manner, the contribution of spontaneous Raman scattering to the read-out noise would be significantly reduced. 

Low optical depth limits the total internal efficiency of the memory. In the explored temperature range, an OD of \numrange[range-phrase=--]{1}{2.5} is reached on the unpumped signal transition. According to the authors of Ref.~\cite{Gorshkov2007}, for $\text{OD} = 5$ (full atomic polarization) the maximal total efficiency achievable with forward retrieval is limited to $<\SI{30}{\percent}$.
Given this, our memory is remarkably efficient.
Considering the magnetic field homogeneity, characterized in \cite{Mottola2023}, a longer cell could be used to increase the OD at constant Rb number density. For a \SI{10}{\milli\meter}-long vapor cell, frequency shifts of at most a few megahertz are expected. By internally routing the light within the cell, longer optical paths can be obtained exploiting the capabilities of wafer-based fabrication techniques \cite{Perez2009,Chutani2015,Nishino2021}.
Furthermore, by designing the transverse profile of the cell to match the interaction region and by applying anti-relaxation coatings to the cell walls, the lifetime of the memory could be extended.

The discrepancy between the measured efficiency and the theoretical expectation can be fully resolved by effects we account for.
The largest contribution is caused by unintentional read-out occurring during storage, constituting nearly a quarter of the achieved end-to-end efficiency. This eliminates the discrepancy within experimental uncertainty. 
In order to address this issue, an investigation into improving the extinction ratio of the control field is ongoing. 
First promising results were obtained by directly switching the TA's operating current. Over 6 orders of magnitude of suppression of the control field are reached almost immediately upon switching \cite{Mottola2023}, which should prevent perturbation of the spinwave during storage.
Additionally, it would be interesting to investigate how the cell's flat form-factor affects the re-emission mode, as it does not appear to remain perfectly matched to the SM fiber mode defined at the input. We also note that for efficient optical pumping its aspect ratio is particularly inconvenient. 

In summary, our implementation of a ground-state quantum memory protocol using laser pulses at the single-photon level in a microfabricated vapor cell opens up a route toward realistic scaling of quantum networks.
Moreover, we have shown that our novel high magnetic field scheme isolates a near-ideal three-level system, even on the $^{87}$Rb D$_2$ line, promising highly efficient and low noise performance in an optimized cell where the necessary optical depths can be reached at lower number densities.
Furthermore, in such a cell, longer memory lifetimes are plausible, as in MEMS atomic clocks ground state coherence times of milliseconds can be reached \cite{Knappe2008}.
Once the atomic state preparation is further improved, interfacing the memory with a compatible single-photon source, e.g., based on downconversion \cite{Buser2022} or a GaAs quantum dot \cite{Zhai2022} will be attempted.
In fact, we note that the presented memory is already built for asynchronous operation, capable of on-demand read-in and read-out. \\

\begin{acknowledgments}
The authors thank Gaetano Mileti for providing us with the vapor cell and Florian Gruet for his help with the characterization measurements concerning the cell's filling. We thank Andrew Horsley for early conceptual contributions and acknowledge financial support from the Swiss National Science Foundation through NCCR QSIT and from the European Union through the Quantum Flagship project macQsimal (Grant No. 820393).
\end{acknowledgments}

\end{document}